\documentclass[12pt]{iopart}
\usepackage{iopams,setstack}
\usepackage{showlabels,showtags}
\usepackage{graphicx,verbatim}
\pdfoutput=1
\def\epsilon{\varepsilon}
\def\hat{\widehat}
\def\tilde{\widetilde}

\newcommand{\ri}{{\rm i}}
\newcommand{\rb}{{\rm b}}
\newcommand{\hphi}{\tilde{\phi}}
\newcommand{\hpsi}{\tilde{\psi}}

\newcommand{\vb}{\boldsymbol{b}}

\newcommand{\ve}{\boldsymbol{e}}

\newcommand{\vf}{\boldsymbol{f}}
\newcommand{\vj}{\boldsymbol{j}}

\newcommand{\vpsi}{\bpsi}
\newcommand{\vphi}{\bphi}

\newcommand{\vnu}{\bnu}

\newcommand{\vu}{\boldsymbol{u}}
\newcommand{\vU}{\boldsymbol{U}}

\newcommand{\vE}{\boldsymbol{E}}
\newcommand{\vJ}{\boldsymbol{J}}
\newcommand{\vv}{\boldsymbol{v}}
\newcommand{\vw}{\boldsymbol{w}}

\newcommand{\vcurl}{\mbox{\bf Curl}}
\newcommand{\vgrad}{\mbox{\bf Grad}}

\newcommand{\vV}{\boldsymbol{V}}

\newcommand{\beq}{\begin{equation}}
\newcommand{\eeq}{\end{equation}}
\begin{document}
\title[Transport current in superconducting films]{Transport current and magnetization problems
for thin type-II superconducting films}

\author{John W. Barrett$^{1}$,  Leonid Prigozhin$^2$ and Vladimir Sokolovsky$^3$}
\address{$^1$Department of
Mathematics, Imperial College London, London SW7 2AZ, UK.}
\address{$^2$Department of Solar Energy and Environmental Physics,
Blaustein Institutes for Desert Research, Ben-Gurion University of
the Negev, Sede Boqer Campus, 84990, Israel.}
\address{$^3$Physics Department, Ben-Gurion University of
the Negev, Beer-Sheva, 84105, Israel}
\eads{\mailto{j.barrett@imperial.ac.uk}, \mailto{leonid@math.bgu.ac.il}, \mailto{sokolovv@bgu.ac.il}}
\date{\small \today}
\begin{abstract}
Thin film magnetization problems in type-II superconductivity are usually formulated in terms of the magnetization function alone, which allows one to compute the sheet current density and the magnetic field but often inhibits computing the electric field in the film. Accounting for the current leads presents an additional difficulty encountered in thin film transport current problems.

We generalize, to the presence of a transport current, the two-variable variational formulation proposed recently for thin film magnetization problems. The formulation, written in terms of the magnetization function and the electric field,  is used as a basis for a new numerical approximation enabling us to solve the magnetization and transport current problems for flat films of arbitrary shapes, including multiply connected films. The advantage of this approach is in its ability to compute accurately all variables of interest, including the electric field, for any value of the power in the power law current-voltage relation characterizing the superconducting material. In the high power limit the critical state model solution is obtained.
\end{abstract}
\maketitle

\section{Introduction}
Numerous applications of thin film type-II superconductors have caused much interest to modeling
their nonlocal electrodynamics and hysteretic response to external magnetic fields and applied transport currents. Mathematical models employed for the macroscopic description of these processes in such superconductors are, usually, eddy current models with a strongly nonlinear current-voltage relation; typically, the relation in the form of a power law \cite{Rhyner} or its high-power limit, the critical state model, is assumed.

The analytical solutions are known for the Bean critical state model \cite{Bean} in the simplest geometrical configurations, such as a thin disk in an external field (the magnetization problem, see \cite{MikhKuz,ClemSanchez}) or an infinite strip (both the magnetization and transport current problems) \cite{BrandtIndenbomForkl,BI,Zeldov}.

Numerical methods for modeling magnetization of flat films of arbitrary shapes have been also developed, see \cite{Brandt95,Brandt95b,SchusterB96,P98,BP2000,NSDVCh,VSGJ07,VSGJ08,VSGJ12}, and have helped to understand those peculiarities of magnetic field penetration and current density distribution in thin films that are absent in disk or strip geometries. These methods were based on the reformulation of the film eddy current problems in terms of one variable, the stream (magnetization) function $g$ introduced for the sheet current density $\vj$ having a zero divergence in the film. The current density was then found as a 2d curl of this function and the magnetic field calculated using the Biot--Savart law.  Accurate computation of the electric field, which determines the local energy dissipation in the film and is often extremely nonuniform, remained, until recently, a difficult problem for models with a power law current voltage relation, if the power is high, and for critical state models.

A new (mixed) variational formulation for thin film magnetization problems written in terms of two variables, the electric field and the magnetization function, has been proposed in \cite{BPSUST} and studied mathematically in \cite{BPfilmmath}. A numerical approximation based on this formulation employed continuous piecewise linear elements for the magnetization function and the Raviart--Thomas element of the lowest order for the electric field; this approximation allowed us to compute also the electric field and was accurate for arbitrary high powers in the current-voltage relation characterizing the superconducting material  \cite{BPSUST,BPfilmmath}.

An additional difficulty in transport current problems is the necessity to deal with the leads supplying this current to the film. Although the solution of these problems is of much practical interest, to the best of our knowledge  they have been solved numerically only for thin films in the Meissner state (\cite{SPD}, see also \cite{VNS}).

In this work we simplify the derivation of the mixed variational formulation \cite{BPSUST} for thin film problems and generalize it to the transport current problems using an approach similar to that in \cite{SPD} to account for the semi-infinite leads, characterized by the same current-voltage relation as the film itself. We also propose a different numerical approximation based upon nonconforming piecewise linear elements for the magnetization function and piecewise constant vectorial elements for the electric field. Such an approximation is much simpler
than that in \cite{BPSUST}, where the Raviart--Thomas element has been employed. Finally, we present an iterative numerical
algorithm, check its accuracy, and present our simulation results for a simply and a multiply connected film. Our new method is applicable to both the magnetization and transport current problems, is simpler to implement and usually faster than that in \cite{BPSUST}, and is also accurate for power law relations with an arbitrary power value.

The  method can also be used for computing the distribution of AC losses in films of arbitrary shapes, studying quench conditions in fault current limiters based on superconducting films of meander or spiral shapes, modeling SQUIDs and the superconducting chips creating magnetic traps for ultra-cold atoms.

\section{Magnetization problems}
Let, in the infinitely thin approximation, the film occupy a bounded 2d domain $\Omega$ lying in the plane $x_3=0$. We assume the film is characterized by the power current voltage relation between the component of the electric field tangential to the film, $\ve$, and the sheet current density
$\vj$:
\beq \ve=e_0\,(|\vj|/j_{\rm c})^{p-1}\vj/j_{\rm c}.\label{plaw}\eeq
Here $e_0$ and the power $p$ are constants, $j_{\rm c}$ is the critical sheet current density, assumed in this work either constant or dependent only on $x=\{x_1,x_2\}\in\Omega$. By $h_3^{\rm e}$ we denote the normal to the film component of the given non-stationary external magnetic field.
The mixed formulation \cite{BPSUST} for the magnetization problem can be more easily derived as follows.

The normal to film component of the total magnetic field can be expressed via the Biot--Savart law,
\begin{eqnarray}
h_3
&=h_3^{\rm e}+\frac{1}{4\pi}\,\mbox{Curl}\int_{\Omega}\frac{\vj(x',t)}{|x-x'|}\,{\rm d}x'\label{BiotS0}\\
&=h_3^{\rm e}+\frac{1}{4\pi}\int_{\Omega}\vcurl'\left(\frac{1}{|x-x'|}\right)\cdot\vj(x',t)\,{\rm d}x', \label{BiotS}
\end{eqnarray}
where  
the 2d  operators
$\mbox{Curl}\,\vf(x):=\partial_{x_1}f_2-\partial_{x_2}f_1$ and
$\vcurl\, u(x):=\left\{\partial_{x_2}u,-\partial_{x_1}u\right\}$.
Using the Faraday law we obtain
\beq\mu_0\,\partial_t h_3=-\mbox{Curl}\,\ve,\label{Far}\eeq
where $\mu_0$ is the permeability of vacuum.
It is convenient to use the transformation, $\ve=R\vv$, where
$$R=\left(\begin{array}{cc}0 &-1\\1 &0\end{array}\right).$$
Substituting (\ref{BiotS0}) into  (\ref{Far}) and taking into account that $\mbox{Curl}\,\ve=\mbox{Div}\,\vv:=\partial_{x_1}v_1+\partial_{x_2}v_2$, we arrive at
\beq
-\frac{1}{\mu_0}\mbox{Div}\,\vv=\partial_th_3^{\rm e}+\frac{1}{4\pi}\mbox{Curl}
\int_{\Omega}\frac{\partial_t\,\vj(x',t)}{|x-x'|}
{\rm d}x'. \label{BF}\eeq

Let us assume first that the domain $\Omega$ is simply connected. Then, since the 2d divergence $\mbox{Div}\,\vj=0$ in $\Omega$, we can introduce a stream (magnetization) function $g$ such that $\vj=\vcurl\,g$ in $\Omega$; see \cite{Brandt95} for a discussion of this function. Let $s$ be the counter-clockwise length parametrization of the domain boundary $\Gamma$ and $\vnu$ the unit exterior normal to this boundary.  Then $j_{\nu}=\partial _{s}g$ on $\Gamma$ and, on setting $g|_{\Gamma}=0$, we satisfy the boundary condition $j_{\nu}=0$ for the magnetization problems.

Let $\phi$ be a smooth enough function in $\Omega$ satisfying $\phi|_{\Gamma}=0$ (we assume $\phi\in H_{00}^{1/2}(\Omega)$, see \cite{BP2000,BPfilmmath}). Multiplying (\ref{BF}) by $\phi$, integrating and using Green's theorem, we arrive at the variational equation
\beq
a(\partial_t g,\phi)-\frac{1}{\mu_0}\left(\vv,\vgrad\,\phi\right)=-(\partial_th_3^{\rm e},\phi),\label{form1}
\eeq
where $\vgrad$ is the 2d gradient, $(\vpsi,\vphi):=\int_{\Omega}\vpsi(x)\cdot\vphi(x)\,{\rm d}x$ and the bilinear form
$$a(\psi,\phi):=\frac{1}{4\pi}\int_{\Omega}\int_{\Omega}\frac{\vgrad\,\psi(x)\cdot\vgrad'\,\phi(x')}
{|x-x'|}\,{\rm d}x\,{\rm d}x'.$$
We now rewrite the constitutive relation (\ref{plaw}) as
$$\vj=j_{\rm c}\left(\frac{|\ve|}{e_0}\right)^{r-1}\frac{\ve}{e_0},$$
where  $r=1/p$. Since $\vj=\vcurl\,g$, $|\vv|=|\ve|$ and $R^{-1}\vcurl=-\vgrad$, this relation takes the form
\beq \vgrad\,g=-j_{\rm c}\left(\frac{|\vv|}{e_0}\right)^{r-1}\frac{\vv}{e_0}.\label{form2}\eeq
To complete the mixed formulation (\ref{form1})--(\ref{form2}) of thin film magnetization problems one needs to specify also the initial condition $g|_{t=0}=g^0$.

Although the formulation (\ref{form1})--(\ref{form2}) has already been derived in \cite{BPSUST}, here we significantly simplified its derivation. Unlike all previous formulations of thin film magnetization problems, this formulation is written in terms of two variables, the magnetization function and the electric field. In the $p\rightarrow\infty$ limit, the solution to this problem tends (see \cite{BPfilmmath}) to the solution of the Bean critical state model.

\section{Transport current problems}
We assume the given transport current $I(t)$ is supplied to a thin superconducting film by means of two semi-infinite superconducting strip leads, see figure 1,
lying in the plane $x_3=0$ as the film itself. A non-zero external magnetic field can also be applied. We also assume that, far away from the film, the distribution of the sheet current density in the leads is not influenced by the film current and is as in a straight infinite strip under the same  conditions (the external magnetic field and transport current). The far-away distributions of sheet current density in the leads can then be found as solutions to one-dimensional problems. In our examples below we use the analytical sheet current density distributions for
infinite strips, which are  known for the Bean model \cite{BI,Zeldov}.

\begin{figure}[h]
\begin{center}
\includegraphics[width=10cm]{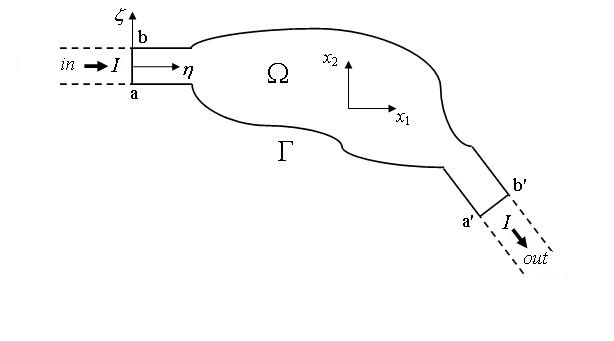}
\end{center}
\caption{Thin film with transport current; dashed lines show the cut-off lead ends.}
\label{Fig1}
\end{figure}
Following \cite{SPD}, we limit our consideration to a bounded 2d domain $\Omega$ which includes the film and sufficiently long parts of the two leads. The influence of the ``cut off" infinite lead ends is accounted for by
adding to the external magnetic field the fields induced by their currents,
$$\tilde{h}^{\rm e}_3=h^{\rm e}_3+h^{\rm in}_{3}+h^{\rm out}_{3},$$
where $h_3^{\rm in}$ and $h_3^{\rm out}$ are the normal to film components of the corresponding induced fields. In the strip-related coordinates $\{\eta,\zeta\}$, see figure 1, for a known sheet current density
$\vj=\{j^{\rm in}_{\eta}(\zeta,t),0\}$ in the semi-infinite strip $(-\infty,0]\times[-w,w]$ we obtain (see \cite{SPD}):
\beq h^{\rm in}_3(\eta,\zeta,t)=\frac{1}{4\pi}\int_{-w}^{w}\frac{j^{\rm in}_{\eta}(\zeta',t)}{\zeta-\zeta'}
\left[1-\frac{\eta}{\sqrt{(\zeta-\zeta')^2+\eta^2}}\right]\,{\rm d}\zeta'\label{hi}\eeq
for $\eta>0$ and, similarly, for $h^{\rm out}_3$.
Contrary to the Meissner state case \cite{SPD}, the sheet current density in this problem is bounded. In particular, for the critical state models $|\vj|\leq j_c$.
Since, for any $\eta>0$ and $\zeta'$ close to $\zeta$,
$$\frac{1}{\zeta-\zeta'}\left[1-\frac{\eta}{\sqrt{(\zeta-\zeta')^2+\eta^2}}\right]\approx \frac{\zeta-\zeta'}{2\eta^2},$$
the integral (\ref{hi}) is not singular. Nevertheless,  careful numerical computation of this field near the cut is desirable; on the cut $\{\eta=0, \ -w\leq \zeta\leq w\}$  the field is equal to half of the magnetic field induced by the infinite strip current \cite{BI} and has logarithmic singularities near the cut ends $\zeta=\pm w$. For $\eta>0$, we used the composite rectangular rule with
a sufficiently large number
of points to compute the integral in (\ref{hi}).

Assuming again the film $\Omega$ is simply connected, we arrive at a similar mixed formulation for the transport current problems: the variational equation (\ref{form1}) with the external field $h_3^{\rm e}$ replaced by $\tilde{h}_3^{\rm e}$, is satisfied for any smooth enough test function $\phi$ such that $\phi|_{\Gamma}=0$, and is supplemented by the nonlinear relation (\ref{form2}). However, the boundary condition for the magnetization function $g$ is now non-homogeneous.

The normal component of the sheet current density $\vj$ on the boundary $\Gamma$ is non-zero only on the two cuts, the parts a--b and a$^{\prime}$--b$^{\prime}$ of $\Gamma$ in figure 1, where we assume $j_{\nu}=-j^{\rm in}_{\eta}$ and $j_{\nu}=+j^{\rm out}_{\eta}$, respectively.
Let us define
\beq g(x,t)=\int_{\Gamma(x_0,x)}j_{\nu}\,{\rm d}s,\label{Psi}\eeq
where $x\in\Gamma$ and $x_0$ is a chosen point on the part a--a$^{\prime}$ of the boundary; this definition makes sense because $\oint_{\Gamma}j_{\nu}\,ds=0$. Obviously, $g=0$  on the part a--a$^{\prime}$ of the boundary and $g=I$ on its part b--b$^{\prime}$.  Since $j_{\nu}=\partial_{s}g$, such a boundary condition ensures the correct values of $j_{\nu}$ on the boundary for transport current problems.

\section{Finite element approximation and numerical algorithm}\label{FEA}
The simplest approximation of the equations (\ref{form1})--(\ref{form2}) would employ continuous piecewise linear elements for the magnetization function and piecewise constant vectorial elements for the electric field. However, although the weak convergence of such an approximation can be proved, our simulations showed that for high power values it yields a ``mosaic structure" of the computed electric field which makes the approximation useless despite its convergence on average (see \cite{BPsandqvi} for a similar phenomenon in another critical state problem).

The finite element approximation, proposed for the magnetization problems in \cite{BPSUST,BPfilmmath}, employed continuous piecewise linear element for the magnetization function $g$ and the lowest order Raviart--Thomas element for the rotated electric field $\vv$.

Here we explore a different approximation for both the magnetization and transport current problems based on the \emph{nonconforming} piecewise linear element for $g$ and the piecewise constant vectorial element for $\vv$, the combination known to be a computationally inexpensive alternative to the Raviart--Thomas element in mixed methods for linear second order elliptic problems, see e.g. \cite{Marini}.

We approximate $\Omega$ by a polygonal domain $\Omega^h$ and denote by ${\cal T}^h$ a regular
partitioning of $\Omega^h$ into triangles $\kappa$ with $h=\max_{\kappa\in {\cal T}^h}\mbox{diam}(\kappa)$ being their maximal size. Here vertices of ${\cal T}^h$ lying on $\Gamma^h$, the boundary of $\Omega^h$,
also lie on $\Gamma$. If $\Omega$ contains subdomains with different critical
current density values, the mesh is fitted in a similar way to the subdomain boundaries.
By ${\cal N}^h$ and ${\cal E}^h$ we denote the sets of nodes and edges of this triangulation,
respectively, with ${\cal N}_{\rm i}^h,\ {\cal E}_{\rm i}^h$  and ${\cal N}_{\rm b}^h,\ {\cal E}_{\rm b}^h$
being the subsets of the internal and boundary nodes and edges, respectively. 

Let us define two finite dimensional spaces of functions defined in $\Omega^h$: $\vU^h$, the space of vectorial functions $\vu(x)$ constant on each triangle, $\vu|_{\kappa}=\vu_{\kappa}\in \mathbb{R}^2$, and the nonconforming linear element space $S^h$ of scalar functions $\psi(x)$  linear on each triangle and continuous at the midpoints $p_e$ of all internal edges $e\in {\cal E}_{\ri}^h$. The functions from $S^h$ are fully determined by their mid-edge values and can be written as
$\psi(x)=\sum_{e\in{\cal E}^h} \,\psi(p_e)\,\phi_e(x)$, where the  basis functions $\phi_e\in S^h$ satisfy
$$\phi_e(p_{e'})=\left\{\begin{array}{ll}1&e'=e,\\0&e'\neq e.\end{array}\right.$$
By $S^h_0\subset S^h$  we denote the subspace of functions which are zero at the boundary edge midpoints.
We extend the definition of the $\vgrad$ operator to discontinuous piecewise linear functions $\psi\in S^h$ by setting
$\vgrad\,\psi:=\vu\in \vU^h$ so that $\vu_{\kappa}=\vgrad\,\psi|_{\kappa}$ for all $\kappa\in {\cal T}^h$.
In addition, let $0=t_0<t_1<...<t_{N-1}<t_N=T$ be a partitioning of $[0,T]$ into possibly variable time steps $\tau_n=t_n-t_{n-1},\ n=1,...,N$.

As the initial condition we take $G^0=\sum_{e\in {\cal E}^h}g^0(p_e)\,\phi_e(x)\in S^h$. For the magnetization problems we define $f^n(x)=-\sum_{e\in {\cal E}^h}h_3^e(p_e,t_n)\,\phi_e(x)$ and set $g(p_e,t_n)=0$
for $e\in{\cal E}_{\rb}^h$.
For problems with transport current we set $f^n(x)=-
\sum_{e\in {\cal E}^h}\tilde{h}_3^e(p_e,t_n)\,\phi_e(x)$ and define $g(p_e,t_n)$ for $e\in{\cal E}_{\rb}^h$ using \eref{Psi} with the boundary    $\Gamma^h$.

Our finite element approximation of these thin film problems is:
\begin{center}
{ Given $G^0\in S^h$, for $n=1,...,N$ find}\\
{$\vV^n\in \vU^h$ and $G^n\in S^h$ such that}\\
{ $G^n(p_e)=g(p_e,t_n)$ for all $e\in {\cal E}_{\rb}^h$,}  \\
\beq a^h\left(\frac{G^n-G^{n-1}}{\tau_n},\phi\right)-\frac{1}{\mu_0}(\vV^n,\vgrad\,\phi)^h=
\left(\frac{f^n-f^{n-1}}{\tau_n},\phi\right)^h\label{Appr1}\eeq
{for any $\phi\in S^h_0$ and, in all $\kappa\in{\cal T}^h$,}
\beq \vgrad\,G^n|_{\kappa}=-j_{\rm c}\left.\left(\frac{|\vV^n|}{e_0}\right)^{r-1}\frac{\vV^n}{e_0}\right|_{\kappa}.\label{Appr2}
\eeq
\end{center}
Here, for $\vu,\, \vw\in\vU^h$,   $$(\vu,\vw)^h:=\sum_{\kappa\in{\cal T}^h}s(\kappa)\,(\vu|_{\kappa}\cdot\vw|_{\kappa}),$$ where $s(\kappa)$ is the area of triangle $\kappa$ and,  for $\psi,\ \phi\in S^h$,
$(\psi,\phi)^h:=\sum_{\kappa}(\psi,\phi)^h_{\kappa}$, where  $$(\psi,\phi)^h_{\kappa}:=\frac{1}{3}\,s(\kappa)\sum^3_{m=1}\psi(p_m^{\kappa})\,\phi(p_m^{\kappa})$$ averages the integrand $\psi\,\phi$ over each triangle $\kappa$ at its mid-edge points $p_m^{\kappa}$, $m=1,2,3$.  Finally, for $\psi,\,\phi \in S^h$,
$$
a^h(\psi,\phi):=\sum_{\kappa\in{\cal T}^h}\sum_{\kappa'\in{\cal T}^h}\left(\vgrad\,\psi|_{\kappa}\cdot\vgrad\,\phi|_{\kappa'}\right)K_{\kappa,\kappa'}$$
with
\beq K_{\kappa,\kappa'}:=\frac{1}{4\pi}\int_{\kappa}\int_{\kappa'}\frac{{\rm d}x\,{\rm d}x'}{|x-x'|}.\label{K}\eeq
Some of the double surface integrals \eref{K} are singular and, for an accurate
approximation of the matrix $K$, we followed the approach in the appendix of \cite{SPD}. In particular, the exact analytical value \cite{Arcioni} was used in the most singular case $\kappa=\kappa'$.

 At each time level we solved the problem \eref{Appr1}--\eref{Appr2} iteratively, setting $\vV^{n,0}=\vV^{n-1}$ and, as in \cite{BPSUST}, approximating the nonlinear term $|\vV^n|^{r-1}\vV^n$ in \eref{Appr2} at the $j^{th}$ iteration by
$$|\bi{V}^{n,j-1}|^{r-1}\,\bi{V}^{n,j-1}+(|\bi{V}^{n,j-1}|_{\epsilon})^{r-1}\,(\bi{V}^{n,j}-\bi{V}^{n,j-1}),$$
where $|\vb|_{\epsilon}=\sqrt{|\vb|^2+\epsilon^2}$ with a small $\epsilon>0$ (in our numerical examples we chose $\epsilon=10^{-10}$). With such an approximation we can rewrite the constitutive relation \eref{Appr2} as \beq\vV^{n,j}=\vV^{n,j-1}-\frac{ |\vV^{n,j-1}|^{r-1}\vV^{n,j-1}+\frac{e_0^r}{j_c}
\,\vgrad\,G^{n,j}}{|\vV^{n,j-1}|_{\epsilon}^{r-1}}.\label{Vj}\eeq
It is sufficient to satisfy \eref{Appr1} for all inner edge basis functions $\phi_e$, $e\in {\cal E}_{\ri}^h$. Substitution of \eref{Vj} into \eref{Appr1} yields, at each iteration, a linear algebraic system with a dense
symmetric positive definite matrix for the unknown $G^{n,j}$ values at the midpoints of all inner edges.

Convergence of these iterations can be accelerated by the over-relaxation: substituting the found solution of the linear system, $G^{n,j}$, into \eref{Vj}, we calculated $\vV^{n,j}$ but then replaced it by
$\alpha\vV^{n,j}+(1-\alpha)\vV^{n,j-1}$ with some $\alpha>1$. In all numerical examples we chose $\alpha=1.8$.

The simplest way to extend these formulation and the numerical method to the case of a multiply connected film is to fill-in the ``holes" of the film by setting $j_c$ in the holes to be very small \cite{P98,BPSUST}. We note, however,
that eddy current models allow one to determine the electric field only inside the conductors. The
approximation of $\ve$ in the holes becomes meaningless in the $j_c\rightarrow 0$ limit and should be disregarded;
only the tangential component of the electric field in the film itself can be found in this model.

\section{Computing the sheet current density and the magnetic field\label{jH}}
Directly, the mixed formulation \eref{form1}--\eref{form2} determines the auxiliary variable $g$ and the electric field $\ve=R\vv$ in the film. Solving the problem numerically, we find the approximations $G^n\in S^h,\ \vE^n\in \vU^h$ to these two variables. Usually, it is also required to find
approximations to the sheet current density $\vj$ and the magnetic field or, at least, its normal to the film component $h_3$, that can be compared to magneto-optical measurement results.

Using the piecewise linear nonconforming approximation of $g$ we can find the approximate
mid-triangle values $G^n(o^{\kappa})=\frac{1}{3}\sum^3_{m=1}G^n(p_m^{\kappa})$; the level contours of $g$ based on these values yield good approximations to the current density streamlines. We found, however, that the  triangle current density values, $\vJ^n=\vcurl \,G^n=-R\,\vgrad\,G^n\in \vU^h$, are, for the obtained approximation of $g$, extremely scattered and cannot be used directly. Nevertheless, the weak (on average) convergence of these values can be shown and a more accurate approximation of $\vj$ can be obtained for the vertex values averaged over all triangles to which the vertex belongs.

A much better piecewise constant approximation of the sheet current density $\vj$ employed in this work is based on a numerical procedure in which the current density is calculated as the curl of a \emph{continuous} piecewise linear approximation, $\tilde{G}^n$, of $G^n$, satisfying the boundary condition at the boundary vertices (see Appendix A).

As in \cite{P98,BPSUST}, the magnetic field induced by the film current was found from the Biot--Savart law \eref{BiotS}.
Here we will limit our consideration to the approximation of the normal to the film component of this field $h_3$ (see, e.g., \cite{SPD} for the approximation of the magnetic field outside the film). At the center $o^{\kappa}$ of a triangle $\kappa\in {\cal T}^h$ we obtain
\begin{eqnarray*}
H_3^n(o^{\kappa})&=\tilde{h}_3^{{\rm e}}(o^{\kappa},t_n)+\frac{1}{4\pi}\sum_{\kappa'\in {\cal T}^h}\vJ^n|_{\kappa'}\cdot\int_{\kappa'}\vcurl'\left(\frac{1}{|o^{\kappa}-x'|}\right){\rm d}x'\\
&=\tilde{h}_3^{{\rm e}}(o^{\kappa},t_n) +\frac{1}{4\pi}\sum_{\kappa'\in {\cal T}^h}(R\vJ^n)|_{\kappa'}\cdot\oint_{\partial\kappa'}\frac{\vnu_{\kappa'}}{|o^{\kappa}-x'|}.
\end{eqnarray*}
Here 
$\vnu_{\kappa'}$ is the outward unit normal to $\partial\kappa'$, the boundary of triangle $\kappa'$, and
the integrals over each triangle edge are approximated using Simpson's quadrature rule.

\section{Simulation results}
The mixed variational formulation \eref{form1}--\eref{form2} was used for numerical solution of  thin film magnetization problems in \cite{BPSUST,BPfilmmath}, where the  Raviart-Thomas element was employed. Our new approximation by nonconforming piecewise linear elements, \eref{Appr1}--\eref{Appr2}, is simpler. To compare the two approaches, we used the same test example as in \cite{BPSUST}: the thin disk magnetization problem having an analytical solution for the Bean model. The  power law model employed here approximates the Bean model if the power is high, see \cite{BPfilmmath}. Both numerical methods were stable for arbitrary high power values; we used $p=10^5$ in this example.  The new method was typically faster and, for the same spatiotemporal mesh, approximated the electric field with much higher accuracy; accuracy of the computed sheet current density was approximately the same for the two methods.
Since, on every iteration of both these methods, a linear system with a dense matrix needs to be solved, the solution becomes time and memory consuming if the  finite element mesh is fine.

Our program, written in Matlab 2012a (64 bit), allowed us to perform simulations on meshes containing up to about eleven thousand elements on a PC with an Intel(R) Core i5-2400 3.1GHz CPU and 8GB RAM; solutions obtained with such or even a significantly cruder mesh are usually sufficiently accurate for practical purposes. However, to further investigate the performance of our method, we ran this program on a more powerful computer with 64GB RAM and 2.0HGz Intel(R) Xeon E5-2620 2x6 CPUs, allowing us to refine the mesh further.

Below we present our simulation results for the more difficult and less studied transport current case and assume for simplicity that no external field is applied.  In these examples we also assume the Bean model with $j_c=1$ and, correspondingly, solve the problem numerically for the power law current voltage relation with a very high power. Our simulations showed that the number of iterations per time step does not
depend strongly either on the power $p$ or on the mesh size.

For the critical state model it can be shown that, if the directions of the electric field in the field penetration zone do not change with time, the current density and the magnetic field can be computed in one time step \cite{BPfilmmath}. Our numerical experiments, starting from the initial state $g^0=0$, confirmed this is true in the case of a monotonically increasing transport current. The electric field is, however, not rate-independent. Hence our numerical strategy for a gradually increasing with time transport current was to obtain a solution for the time $t$ in two steps: a large time step
$\tau_1$ was followed by a small time step $\tau_2=t-\tau_1$. The computed electric field can then be regarded as an approximation to the mean electric field during the second time step or, as we did in this work,  to the field at the time $t-0.5\,\tau_2$.

It is convenient to use dimensionless variables,
\begin{eqnarray*}&\hat{x}=\frac{{x}}{{L}},\quad \hat{t}= \frac{{t}}{{t}_0},\quad
\hat{\bi{e}}=\frac{{\bi{e}}}{{e}_0},\quad \hat{\bi{j}}=\frac{{\bi{j}}}{{j}_{\rm c}},\\
&\hat{h}_3=\frac{h_3}{{j}_{\rm c}},\quad \hat{g}=\frac{{g}}{{j}_{\rm c}{L}},\quad \hat{I}=\frac{{I}}{{j}_{\rm c}{L}},
\end{eqnarray*}
where ${L}$ is the length scale characterizing the film size and  ${t}_0= {\mu}_0\,{j}_{\rm c}\,{L}/{e}_0$
is the time scale.
Below we use the dimensionless quantities but, for simplicity, omit the sign `` $\hat{ }\,$ ".

Let a semi-infinite current lead of width $2{w}$ be described by the critical state Bean model with the same sheet critical current density as the film itself. Far from the film the dimensionless sheet current density in the lead  is as in an infinite strip \cite{BI}:
\beq
j_{\eta}(\zeta,t)=\left\{\begin{array}{lc}\frac{2}{\pi}\arctan\sqrt{\frac{w^2-b^2}{b^2-\zeta^2} } & |\zeta|<b,\\
1 & b\leq|\zeta|\leq w,\end{array}\right.
\label{an}\eeq
where  $b(t)=w\sqrt{1-(I(t)/2w)^2}$ and ${\eta,\zeta}$ are the dimensionless strip-related coordinates (see figure 1). We set $g=0$ for $\zeta=-w$ and,  integrating
\eref{an} to find $g=\int_{-w}^{\zeta}j_{\eta}(\zeta',t){\rm d} \zeta'$ on the cut $\eta=0,-w\leq \zeta\leq w$,  obtain:

\beq
g(\zeta,t)=\left\{\begin{array}{lc}w+\zeta & -w\leq \zeta\leq -b,\\
\frac{I}{2}+ \frac{2}{\pi}\,{\cal G}(\zeta,t) & |\zeta|<b,\\
I+\zeta-w &b\leq \zeta\leq w,
\end{array}\right.\label{g_an}\eeq
where
\begin{eqnarray*}{\cal G}(\zeta,t)=\sqrt{w^2-b^2}
\arctan\left(\frac{\zeta}{\sqrt{b^2-\zeta^2}}\right)+\\\quad\quad\zeta \arctan\left(\sqrt{ \frac{w^2-b^2}{b^2-\zeta^2} }\right)
-w\arctan\left(\frac{\zeta}{w}\sqrt{\frac{w^2-b^2}{b^2-\zeta^2}}\right).
\end{eqnarray*}
The expression \eref{g_an} sets the boundary condition for the magnetization function on the lead cuts.

As a test for our computations, let us consider the infinitely long strip $(-\infty,\infty)\times[-0.5,0.5]$, assume $g^0=0$, and apply the dimensionless transport current $I(t)=t$. The Bean model sheet current density distribution is given by \eref{an} with $w=0.5$. The normal to the strip component of the magnetic field (see \cite{BI}) is
$$h_3=\left\{\begin{array}{ll} 0 & |\zeta|< b,\\
\frac{1}{\pi}\,\mbox{sign}(\zeta)\,\mbox{arctanh}\sqrt{\frac{\zeta^2-b^2}{0.5^2-b^2}}& b\le |\zeta|<0.5,
\end{array}\right.$$
and, using the Faraday law, we obtain $\ve=\{e_{\eta},0\}$ with
$$e_{\eta}=\left\{\begin{array}{ll} 0 & |\zeta|< b,\\
\frac{1}{2\pi}\ln\left(\frac{|\zeta|}{b}+\sqrt{\frac{\zeta^2}{b^2}-1}\right)& b\le |\zeta|<0.5.
\end{array}\right.$$
To this analytical solution we compared a numerical one computed for the square film  $\Omega=[-0.5,0.5]\times[-0.5,0.5]$  cut out of the strip and characterized by the power law current voltage relation with $p=10^{6}$. We 
found the approximate solution for $t=0.75$ in two time steps, $\tau_1=0.7$ and $\tau_2=0.05$. For the uniform $50\times 50$ mesh (5,000 elements) the relative errors (in the $L^1$ norm) were $\delta(\vj)=1.0\%$, $\delta(\ve)=4.7\%$ and $\delta(h_3)=2.9\%$; for the $100\times 100$ mesh (20,000 elements) the errors were $\delta(\vj)=0.5\%$, $\delta(\ve)=3.3\%$ and $\delta(h_3)=1.5\%$.
The computation times, not including computing the matrix $K$, see (\ref{K}), were, correspondingly, 17 minutes and 16 hours. Although the right-hand side of (\ref{Appr1}) involves $f^n$, which is singular
in the corners of $\Omega$, the stated quadrature involving $(\cdot,\cdot)^h$ was adequate on these meshes
in that more sophisticated quadrature rules resulted in almost identical results.

We present here also results of our numerical simulations for two other films.
In both cases we assumed again the transport current $I(t)=t$ and, as above, used the time steps $\tau_1=0.7$ and $\tau_2=0.05$ and the power $p=10^6$.

\begin{figure}[h]
\begin{center}
\includegraphics[width=7cm]{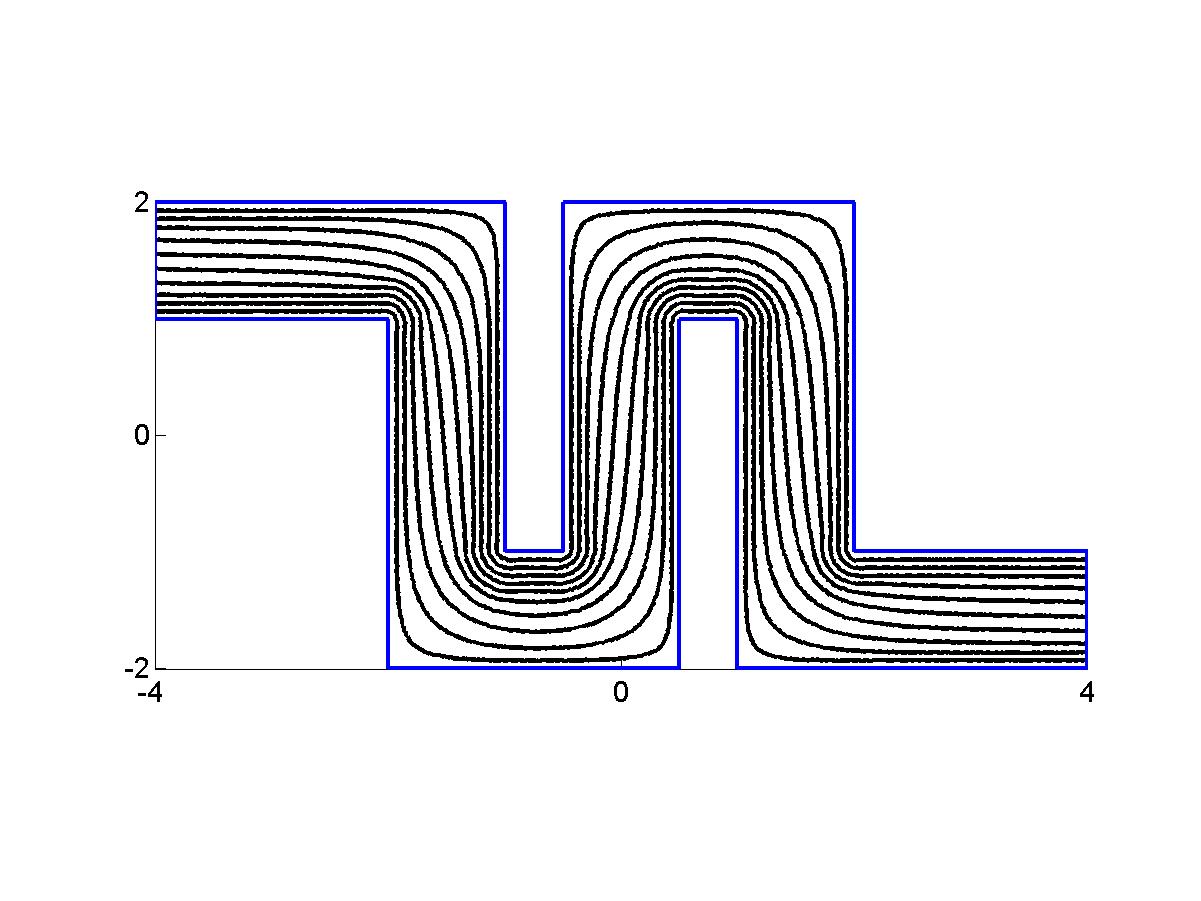}\hspace{.3cm}\includegraphics[width=7cm,height=8cm]{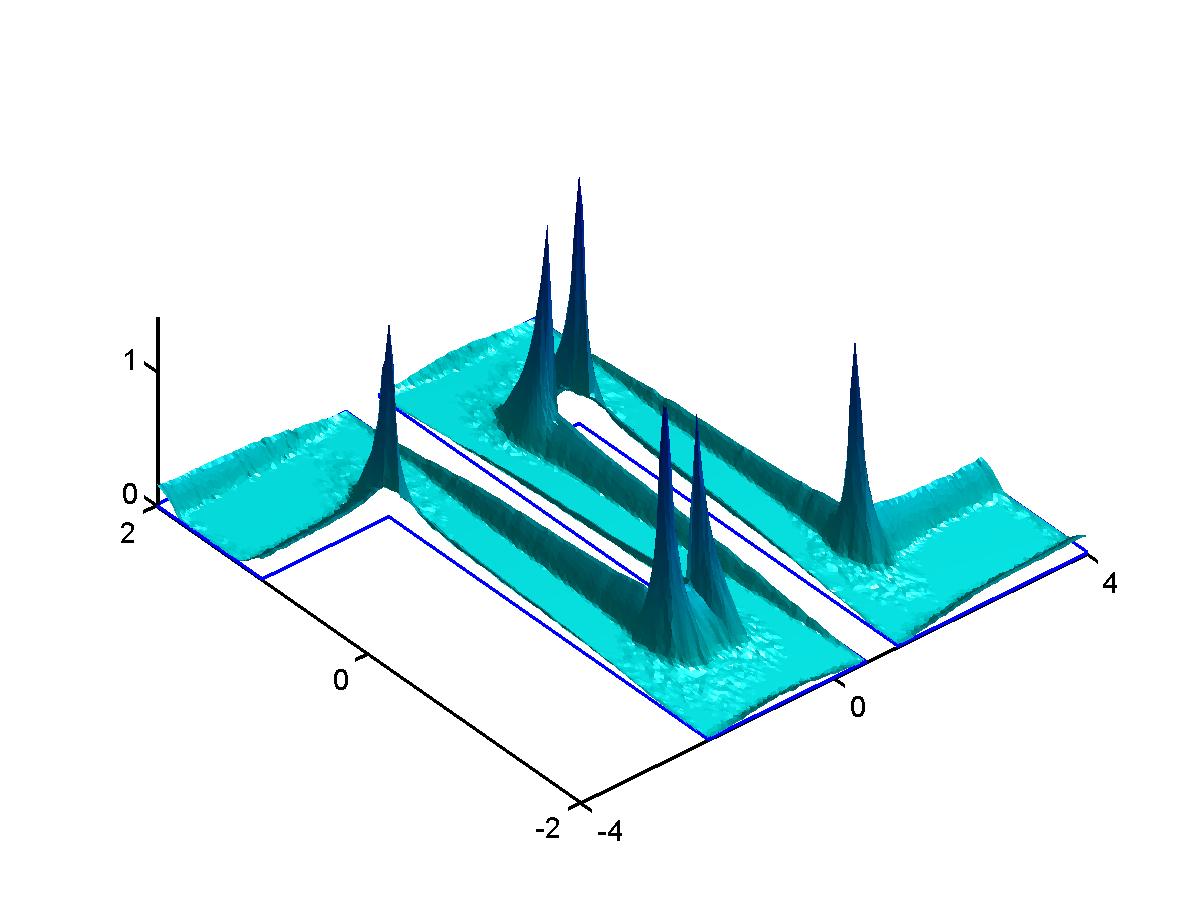}\\
\includegraphics[width=7cm,height=8cm]{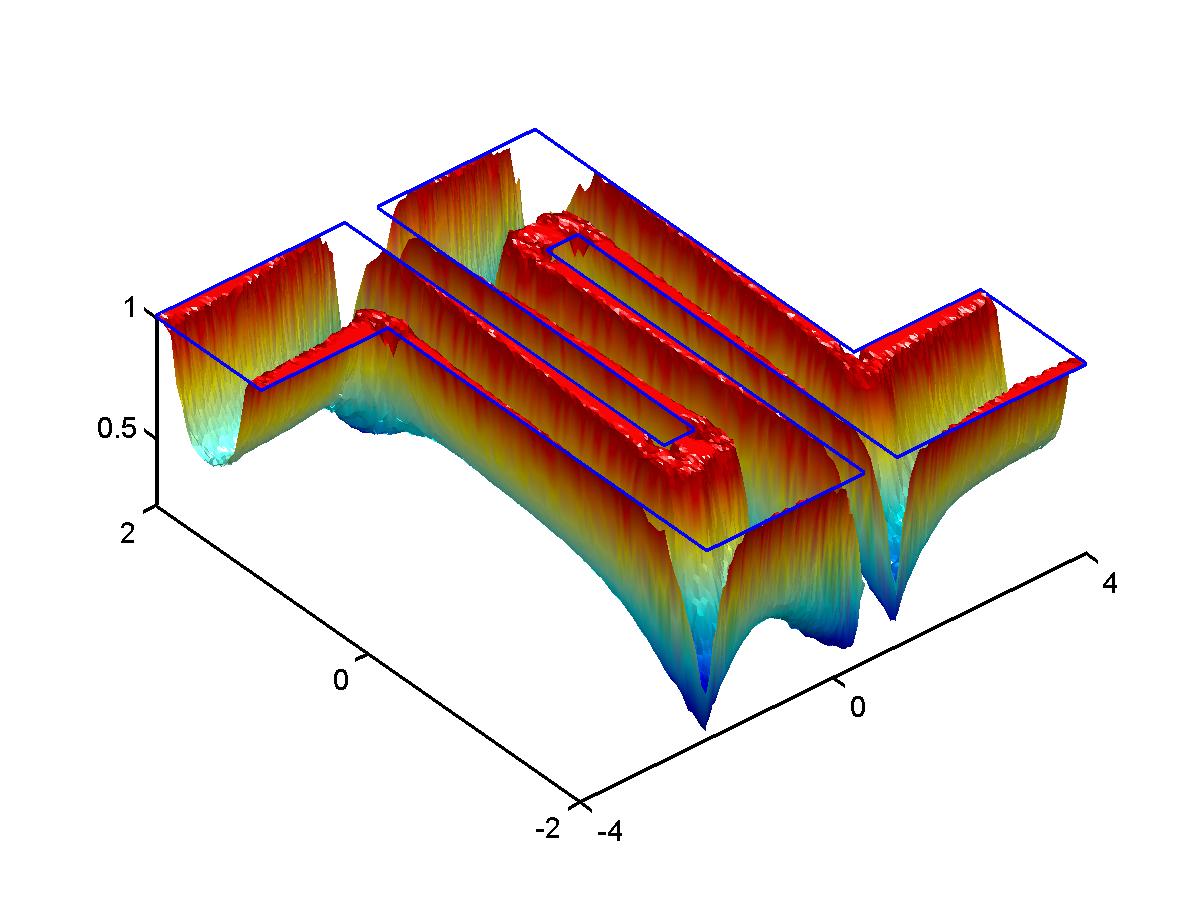}\hspace{.3cm}\includegraphics[width=7cm,height=8cm]{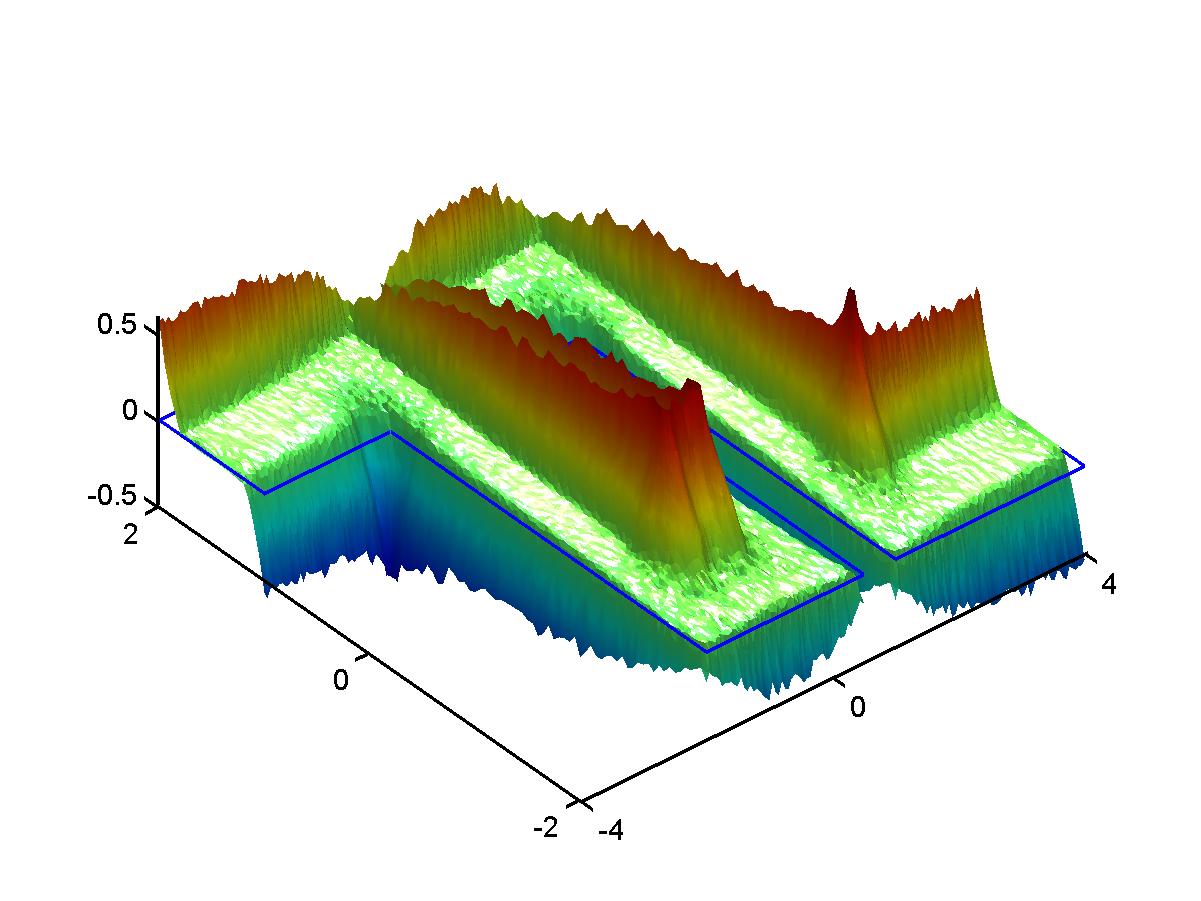}
\end{center}
\caption{Transport current in a meander-shaped film, simulation results for $I(t)=t$ at $t=0.75$. Left: the current stream lines plotted as level contours of the magnetization function  (top) and the modulus of the sheet current density (bottom). Right: the modulus of the tangential to the film component of the electric field (top) and the normal to the film component of the magnetic field (bottom). The blue lines indicate the film boundary.
}
\label{fig_meander}

\end{figure}
The first film (figure \ref{fig_meander}) resembles the meander-shaped films employed in fault current limiters \cite{Otabe}; our mesh in this example contained about 23 thousand elements.  
Near to the lead cuts the solution remains close to that for the infinite strip; increasing the length of the remaining lead ends does not change the solution in the film.  The sheet current density drops down near the convex corners of the film, while an extended critical current density region is observed near the concave corners. The electric field, non-zero (up to a small computational error) only in the critical current density zone, becomes singular in the concave corners of the film. The normal to the film magnetic field component is also zero in the region of subcritical current and penetrates further into the film in the vicinity of the concave corners. This field is singular on the film boundary. Solutions to magnetization problems demonstrate similar behaviour \cite{BPSUST}.

\begin{figure}[h]
\begin{center}
\includegraphics[width=7cm]{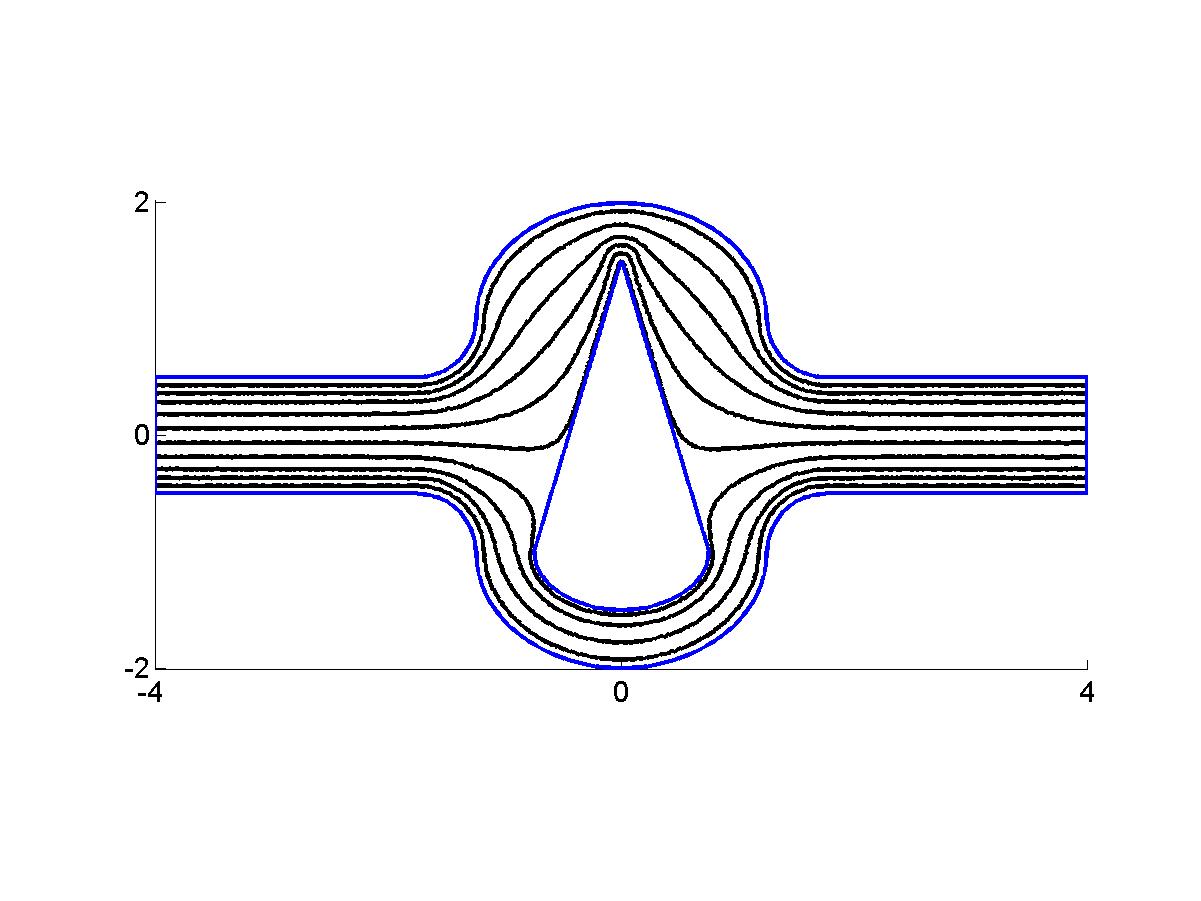}\hspace{.3cm}\includegraphics[width=7cm,height=8cm]{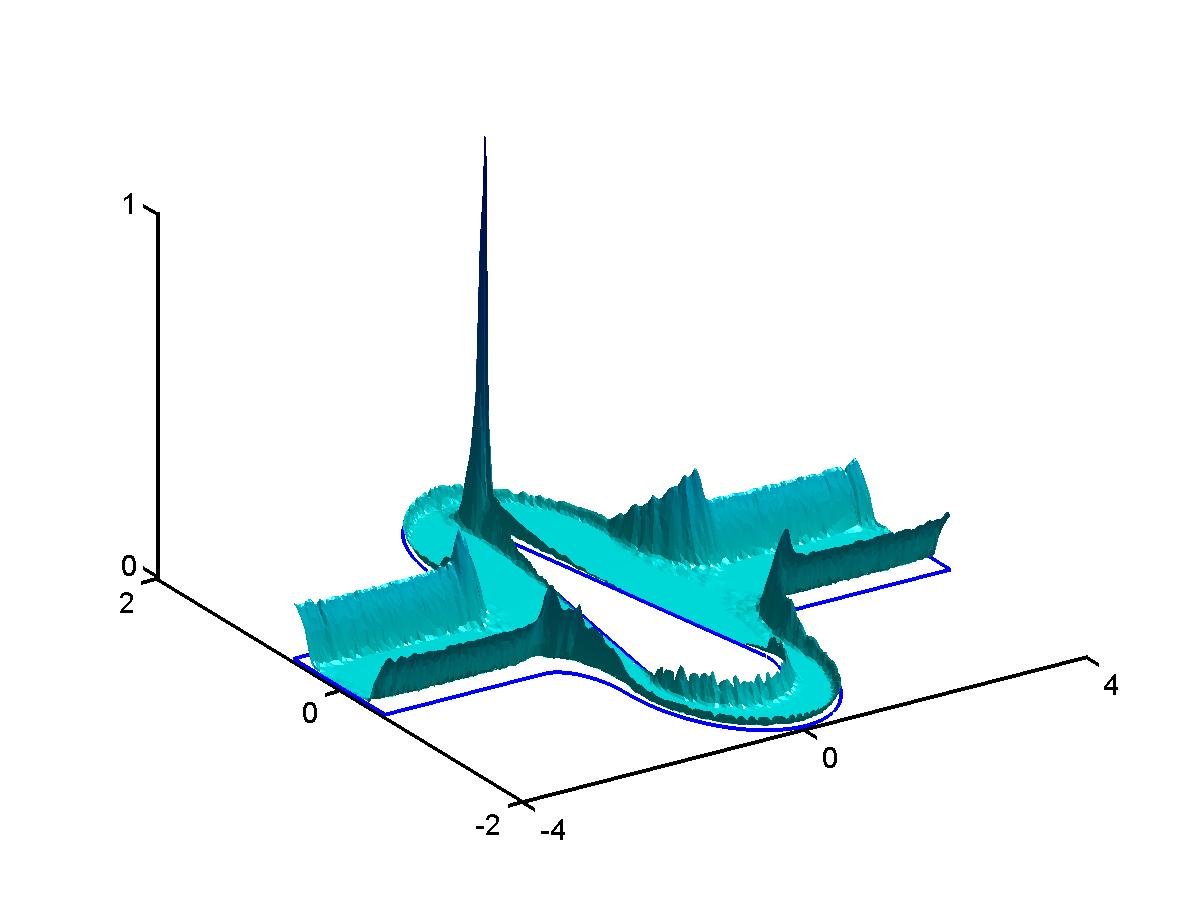}\\
\includegraphics[width=7cm,height=8cm]{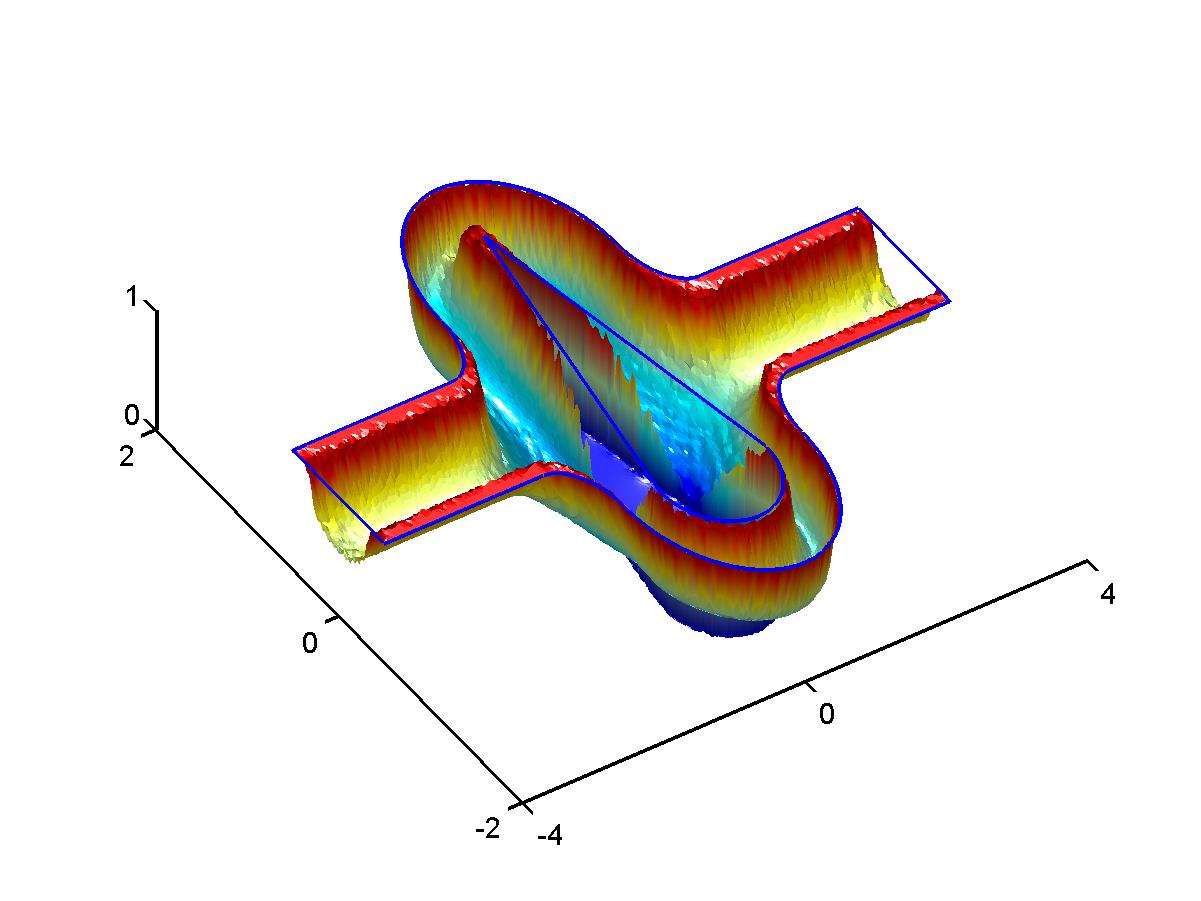}\hspace{.3cm}\includegraphics[width=7cm,height=8cm]{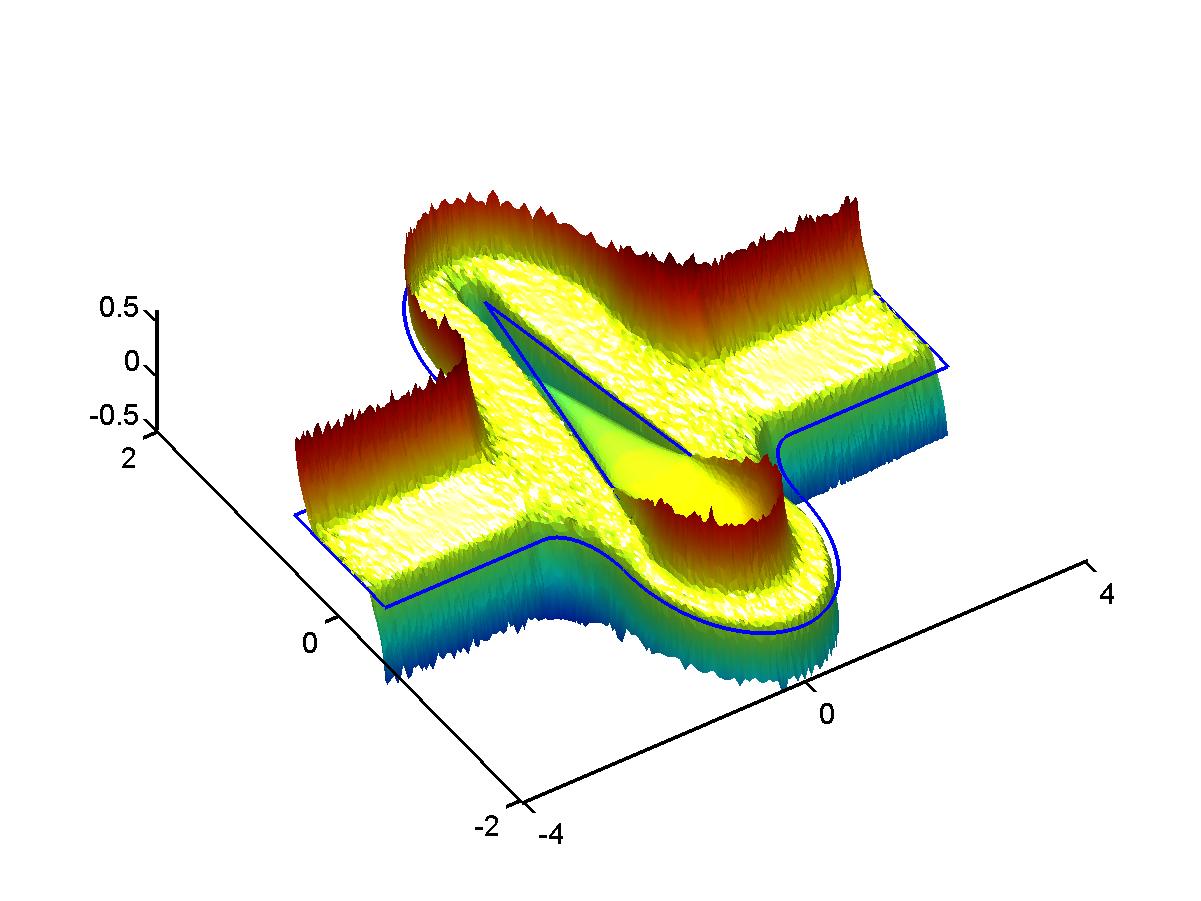}
\end{center}
\caption{As in figure \ref{fig_meander}, but for a multiply connected film.
}
\label{fig_hole}
\end{figure}

As our second example we considered a film with a hole (figure \ref{fig_hole}). Setting in the hole the dimensionless critical sheet current density $j_c=10^{-3}$, we solved the problem in the whole domain keeping $j_c=1$ in the film itself. In this case the mesh contained about 19 thousand elements. 
The magnetic field $h_3$ penetrates into the critical sheet current density region and, differently to the case of cylindrical superconductors, is not uniform in the hole of the thin film. The nonzero electric field is observed near the rounded concave part of the hole boundary and is  singular in the vicinity of the hole corner. In the hole itself the computed electric field $\ve$ is disregarded and is not shown in this figure: since the eddy current model employed does not determine the electric field in insulators, the field in the hole computed in the $j_c\rightarrow 0$ limit hardly makes any sense.

\clearpage

\section{Conclusion} Numerical methods for thin type-II superconducting film magnetization problems, based on the problem formulation in terms of the magnetization function alone, are often unable to ensure accurate calculation of the electric field distribution in a film. Here we employed the two-variable formulation proposed for magnetization problems in \cite{BPSUST}. We extended this formulation to transport current problems, which have been solved previously only for films in the Meissner state \cite{SPD,VNS} and, analytically, for an infinite strip \cite{BI,Zeldov}. We also derived a new finite element approximation, simpler but not less efficient than in \cite{BPSUST}, and estimated the numerical approximation error using test problems with analytical solutions. In both the magnetization and transport current problems all variables of interest were found with good accuracy, which makes it possible to compute also the AC loss distribution in a thin flat film of arbitrary shape.

The results of  numerical simulations are in this work presented for the more difficult transport current problems. Although we limited our consideration to the Bean critical state model, which we approximated by the power current-voltage relation with a very high power, the method is applicable for any power value in the power law typically employed for characterizing type-II superconductors. The method can be used also if both the external magnetic field and the transport current are applied simultaneously; its extension to the case of a field dependent critical sheet current density (Kim-like models) is also straightforward. In these cases the sheet current density in the straight leads far from the film should be found numerically, as a solution to a 1d problem.

\appendix
\section{Reconstruction of the current density}\setcounter{section}{1}
As noted above (see section \ref{FEA}), the simplest finite element approximation of \eref{form1}--\eref{form2}, based on the \textit{continuous} piecewise linear element for $g$ and the piecewise constant vectorial element for $\vv$,
yields poor results for $\vv$ and should be rejected. A simple but accurate approximation of both variables, $G^n$ and $\vV^n$, proposed in this work, approximates the magnetization function $g$ by means of the \textit{nonconforming} piecewise linear element. It turns out, however, that in this case the sheet current density, computed directly as $\vJ^n=\vcurl\,G^n$, is not accurate. A good approximation of this variable can be computed using the following reconstruction procedure, which is efficient and has no influence on the total computation time.

Let us define the space of {continuous} piecewise linear functions, $$\tilde{S}^h=\{\tilde{\psi}\in C(\Omega^h)\ :\ \tilde{\psi}|_{\kappa} \mbox{\ is linear for all\ } \kappa\in {\cal T}^h\},$$ and its subspace $\tilde{S}^h_0$ of functions which are zero on the boundary of $\Omega^h$. The functions $\tilde{\psi}\in \tilde{S}^h$ are determined by their vertex values, $\hpsi(x)=\sum_{\sigma\in {\cal N}^h}\hpsi(\sigma)\,\hphi_\sigma(x)$, where $\hphi_\sigma \in \tilde{S}^h$ is the standard basis function equal to one at vertex $\sigma$ and to zero at all other vertices.

We compute our approximation to the sheet current density
$\vj$ as $$\vJ^n=-R\,\vgrad\,\tilde{G}^n\in \vU^h,$$ where $\tilde{G}^n\in \tilde{S}^h$ is a continuous piecewise linear  approximation of $G^n$ minimizing
$$\sum_{\kappa\in{\cal T}^h}s(\kappa)\left|\vgrad\,G^n|_{\kappa}-\vgrad\,\tilde{G}^n|_{\kappa}\right|^2$$
over the set of $\tilde{S}^h$ functions satisfying $\tilde{G}^n(\sigma)=g(\sigma,t_n)$ at all boundary vertices $\sigma\in {\cal N}_{\rb}^h$. Values of $\tilde{G}^n$ at the inner vertices can be determined from the equivalent equation
\beq\left(\vgrad\,\tilde{G}^n,\vgrad\,\hpsi\right)^h=
\left(\vgrad\,{G}^n,\vgrad\,\hpsi\right)^h,\label{Lapl}\eeq
which should be satisfied for all $\hpsi\in \tilde{S}_0^h$ and may be considered an approximation of the Poisson equation on the mesh ${\cal T}^h$. It is sufficient to satisfy \eref{Lapl} for the functions $\hphi_\sigma$ for all $\sigma\in {\cal N}_{\ri}^h$; this makes it a linear algebraic system with a sparse symmetric positive definite
matrix.


\begin{thebibliography}{99}

\bibitem{Rhyner} Rhyner J 1993 Magnetic properties and AC-losses of superconductors with
power law current-voltage characteristics \textit{Physica} C \textbf{212}  292--300
\bibitem{Bean}
Bean C P 1964  Magnetization of high-field superconductors \textit{Rev. Mod. Phys.} {\bf 36} 31--39
\bibitem{MikhKuz}
Mikheenko P N and Kuzovlev Yu E 1993 Inductance measurements of HTSC films with
high critical currents \textit{Physica} C {\bf 204} 229--236
\bibitem{ClemSanchez} Clem J R and Sanchez A 1994
Hysteretic losses and susceptibility of thin superconducting
disks \textit{Phys. Rev.} B \textbf{50} 9355--9362
\bibitem{BrandtIndenbomForkl}
Brandt E H, Indenbom M V and Forkl A 1993 Type-II superconducting strip in
perpendicular magnetic field
{\em Europhys. Lett.} {\bf 22}, 735--740
\bibitem{BI} Brandt E H and Indenbom M 1993 Type-II-superconductor strip with current
in a perpendicular magnetic field \textit{Phys Rev} B \textbf{48} 12893-12906
\bibitem{Zeldov}
Zeldov E, Clem J R, McElfresh M and Darwin M 1994 Magnetization and transport currents in thin superconducting strips
{\em Phys. Rev.} B {\bf 49}, 9802--9822
\bibitem{Brandt95} Brandt E H 1995 Square and rectangular thin superconductors
in a transverse magnetic field \textit{Phys. Rev. Lett.} \textbf{74} 3025--3028
\bibitem{Brandt95b} Brandt E H 1995 Electric field in superconductors with rectangular cross section
\textit{Phys. Rev.} B \textbf{52} 15442--15457
\bibitem{SchusterB96} Schuster Th, Kuhn H and Brandt E H 1996 Flux penetration into flat
superconductors of arbitrary shape: Patterns of magnetic and electric fields and current
\textit{Phys. Rev.} B \textbf{54} 3514--3524
\bibitem{P98} Prigozhin L 1998 Solution of thin film magnetization problems in
type-II superconductivity \textit{J. Comp. Phys.} \textbf{144} 180--193
\bibitem{BP2000} Barrett J W and Prigozhin L 2000 Bean's critical-state model as the $p\rightarrow\infty$ limit of an evolutionary $p-$Laplacian equation \emph{Nonlinear Analysis} \textbf{42} 977--993
\bibitem{NSDVCh} Navau C, Sanchez A, Del-Valle N and Chen D-X 2008
Alternating current susceptibility calculations for thin-film superconductors with regions of different critical-current densities \textit{J. Appl. Phys.} \textbf{103} 113907
\bibitem{VSGJ07} Vestgarden J I, Shantsev D V, Galperin Y M and Johansen T H 2007
   Flux penetration in a superconducting strip with an edge indentation
   \textit{Phys. Rev.} B \textbf{76} 174509
\bibitem{VSGJ08} Vestgarden J I, Shantsev D V, Galperin Y M and Johansen T H 2008
   Flux distribution in superconducting films with holes \textit{Phys. Rev.} B \textbf{77} 014521
\bibitem{VSGJ12}  Vestgarden J I, Yurchenko V V, Wordenweber R and Johansen T H 2012
  Magnetic flux guidance by micrometric antidot arrays in superconducting films
  \textit{Phys. Rev.} B \textbf{85} 014516
\bibitem{BPSUST} Barrett J W and Prigozhin L 2012 Electric field formulation for thin film magnetization problems  \textit{Supercond. Science and  Technology} \textbf{25} 104002
\bibitem{BPfilmmath} Barrett J W and Prigozhin L 2013 Existence and approximation of a mixed formulation for thin film magnetization problems in superconductivity (\textit{submitted})
\bibitem{SPD} Sokolovsky V, Prigozhin L and Dikovsky V 2010 Meissner transport current in flat films of arbitrary shape and a magnetic trap for cold atoms \textit{Supercond. Science and  Technology} \textbf{23} 065003
\bibitem{VNS} Via G, Navau C and Sanchez A 2013 Magnetic and transport currents in thin film superconductors of arbitrary shape within the London approximation \textit{J. Applied Phys.} \textbf{113} 093905
\bibitem{BPsandqvi} Barrett J W and Prigozhin L 2012 A quasi-variational inequality problem arising in the modeling of growing sandpiles \textit{ESAIM: Mathematical Modelling and Numerical Analysis} (\emph{to appear})
\bibitem{Marini} Marini L D 1985 An inexpensive method for the evaluation of the solution of the lowest order Raviart--Thomas mixed method \textit{SIAM J. Numer. Anal.} \textbf{22} 493--496
\bibitem{Arcioni} Arcioni P, Bressan M and Perregrini L 1997
On the evaluation of the double surface
integrals arising in the application of the boundary integral method to 3-D problems
\textit{IEEE Trans. on Microwave Theory and Techn.} \textbf{45}  436-439
\bibitem{Otabe} Otabe E S, Endo T, Matsushita T and Morita M 2001 AC current loss of a meander-shaped QMG current limiting device \textit{Physica C} \textbf{357-360} 878-881

\end{thebibliography}
\end{document}